# Research collaboration and productivity: is there correlation?[1]

Giovanni Abramo[a,b,*], Ciriaco Andrea D'Angelo[a], Flavia Di Costa[a]

[a] *Laboratory for Studies of Research and Technology Transfer, School of Engineering, Department of Management, University of Rome "Tor Vergata", Italy*

[b] *Italian Research Council*

**Abstract**

The incidence of extramural collaboration in academic research activities is increasing as a result of various factors. These factors include policy measures aimed at fostering partnership and networking among the various components of the research system, policies which are in turn justified by the idea that knowledge sharing could increase the effectiveness of the system. Over the last two decades, the scientific community has also stepped up activities to assess the actual impact of collaboration intensity on the performance of research systems.

This study draws on a number of empirical analyses, with the intention of measuring the effects of extramural collaboration on research performance and, indirectly, verifying the legitimacy of policies that support this type of collaboration. The analysis focuses on the Italian academic research system. The aim of the work is to assess the level of correlation, at institutional level, between scientific productivity and collaboration intensity as a whole, both internationally and with private organizations. This will be carried out using a bibliometric type of approach, which equates collaboration with the co-authorship of scientific publications.

**Keywords**

*Research collaboration, productivity, universities, bibliometrics*

* Corresponding author, Dipartimento di Ingegneria dell'Impresa, Università degli Studi di Roma "Tor Vergata", Via del Politecnico 1, 00133 Roma - ITALY, tel. +39 06 72597362, fax +39 06 72597305, abramo@disp.uniroma2.it

# 1. Introduction

The use of collaboration has increased and gained in importance in the domain of scientific research over the last few decades. Various factors are responsible for this, including the growing specialization of science, the complexity of investigated problems and the increasing costs of scientific equipment needed to perform experiments. Other factors in favor of increasing collaboration cannot be ignored: results of easier access to public financing; aspirations for greater prestige and visibility resulting from collaboration with renowned research groups; and opportunities to attain higher productivity (Lee and Bozeman, 2005). Furthermore, innovations in information and communication technologies and a general decline in transportation costs (Katz and Martin, 1997) have certainly removed some of the barriers to collaboration and eased the impact of what is known as the "proximity effect", whereby collaboration intensity is inversely proportional to the distance between the players at stake.

In policy, there is now a well-established trend of using specific measures to foster scientific collaboration at both local and transnational levels, since knowledge sharing among researchers is believed to be conducive to a significant increase in research effectiveness, just as specialization generally obtains increases in productive efficiency (Adams et al., 2005).

European Union research policies have acknowledged and supported the creation of networks as essential tools for sharing knowledge and promoting innovation, towards the achievement of specific goals. These policies have included the overall Framework Programmes for research and development. In 2004, European Commission Communication 353 further defined six important objectives, with a view to intensifying the impact of its action. The second of those objectives ("creating European centres of excellence through collaboration between laboratories") includes specific programmes supporting transnational collaboration among research centers, universities and companies, with the intention of significant impacts on the quality of research in Europe and the dissemination of knowledge and research results. Finally, in its Green Paper "New Perspectives for the European Research Area" (2007), the European Commission listed the basic requisites for the full development of a European Research Area: these included sharing of resources, instruments and knowledge between the public and the industrial systems, and also among public organizations in different member States.

While guiding policies at the supra-national and national levels were being developed and implemented in recent years, the scientific community intensified its efforts to assess, among other points, the real impact of collaboration intensity on the performance of research systems. General studies in the academic world have shown that collaborations contribute to scientific productivity and that, as a consequence, national research policies should focus on fostering collaboration (Landry et al. 1996; Lee and Bozeman 2005). Other studies examined individual types of collaboration. In particular, international collaboration produces real and remarkable results in the scientific performance of research groups (Van Raan, 1998; Martin-Sempere et al., 2002; Barjak and Robinson, 2007). As regards collaboration between universities and enterprises, a study of the electronics sector by Balconi and Laboranti (2006) also showed that researchers who are co-inventors of patents with private companies have a significantly higher scientific performance than their colleagues.

The study of phenomena related to scientific collaboration is usually carried out using one of two methodologies. The former, known as qualitative methodology, is aimed at investigating the factors which motivate collaboration and the dynamics which underlie it. On the other hand, quantitative methods are used to map and measure collaboration activities: these include the bibliometric analysis of co-authorship of scientific papers published in international journals and indexed in specialist databases.

This study draws on a number of empirical analyses intended to measure the effects of extramural collaboration on research performance and, indirectly, to verify the legitimacy of policies that support such collaboration. Our analysis covers the Italian academic scientific system, using a bibliometric-type approach in which collaboration and co-authorship of scientific publications are treated on a par, and is aimed at assessing the impact of collaboration intensity on scientific productivity. In this paper, in particular, the following questions are considered:

- Is there a correlation between a university's extramural collaboration intensity and its scientific performance?
- Is international openness, measured in terms of collaboration with foreign organizations, correlated to a university's scientific performance?
- Is collaboration with domestic companies related to the scientific performance of research groups?

While looking for answers to these questions, we will also try to establish what dimensions of research performance are actually connected with the phenomenon of collaboration. The research questions could be rephrased as follows: "Is the quality level of a university's research a predictor for the extramural collaboration intensity of its scientific staff?" A possible virtuous cycle is assumed, whereby more external collaboration produces a better research performance, and a higher scientific reputation brings about an increase in collaboration demand from outside parties.

Compared to the state of the art, the distinctive feature of this study is its comprehensiveness. Collaborations are studied in general and under their individual forms: universities-universities; university-public research institutions; universities-companies; universities-foreign organizations. The study covers the whole universe of the Italian universities rather than any type of sample selected from the universe.

In the remainder of this paper, Section 2 reviews the literature on the topic at hand , while Section 3 describes the methodology, in terms of domain of investigation and data set. Section 4 presents the findings obtained, with reference to the proposed research questions, and Section 5 provides the concluding remarks of the authors.

## 2. Research productivity and collaboration intensity: literature review

The assessment of research productivity has gained increasing importance among scholars and research policy-makers since the 1970's. The topic is a difficult one, due to the multidimensional character of the function of scientific knowledge production, and has been approached from different perspectives, one of the objectives being to identify the determinants and their impact on the performance of individuals and institutions.

Various studies over time have made it possible to identify a number of attributes associated with scientific productivity. These may be roughly divided into three categories: personal attributes (researcher sex, age, education, etc.), institutional and departmental attributes (characteristics of the institution, size of faculty, technology and instrumental infrastructures available, etc.) and environmental attributes (labor policies, public and private funds available, students available to support the research, etc.). More recently, Dundar et al. (1998) studied the productivity of American universities during the period 1988-1992, using an econometric model based on a number of variables, mainly of an institutional and environmental nature. The authors verified a different scientific prolificacy among the various disciplines and observed that faculty size was an important factor in individual productivity, with larger faculties offering researchers better opportunities for collaboration.

A similar study by Ramsden (1994) looked into the Australian university system. Its results broadly confirmed Fox's (1983, 1992a, 1992b) conclusions that high levels of scientific productivity result from the combination of personal and environmental factors.

If the single determinants of productivity are considered, scientific collaboration is among those unanimously recognized as exerting a significant influence on the performance of individual researchers and institutions, in terms of both effectiveness and efficiency. So much so that it has become a cornerstone in research policies at national and supranational level.

Lee and Bozeman (2005) attempted to evaluate the degree to which collaboration among scientists influences scientific productivity, as measured in terms of publications. A sample of American university researchers was surveyed. The results showed that the number of collaborating researchers is the strongest predictor of productivity and that the positive correlation between collaboration and productivity is adequately robust.

Adams et al. (2005) studied the effects of the size of research teams, again within the American university system. Based on the number of authors mentioned in each publication, they calculated the number of internal and international collaborations during the period 1981-1999. As it emerged, scientific production grows as the team becomes larger. It was therefore concluded that increases in scientific productivity result directly from a greater division of labor within larger research groups.

Landry et al. (1996) studied scientific collaborations involving university researchers in Quebec, Canada. The data, collected through a survey and analyzed with an econometric model, showed that collaboration intensity influences productivity to a varying degree, depending on geographical proximity and field of speculation. The data, however, confirmed that collaborations generally contribute to scientific productivity, and therefore national research policies should aim to foster collaborations.

Most studies are limited in scope to a small number of disciplinary sectors. For example, Bordons et al. (1996) analyzed the influence of collaborations on scientific performance for three sectors within the biomedical area: neurosciences, gastroenterology and the area concerning cardiovascular systems. The bibliometric analysis used a number of indicators and showed that international and intramural collaborations are positively correlated with the productivity of individual authors, as collaborations give scientists the opportunity to work on different projects at the same time. Furthermore, more applied types of research (e.g. gastroenterology) are marked by greater collaboration at the national level, whereas basic research (e.g. neurosciences) are rather associated with international collaboration, which ensures greater visibility and the opportunity for publication in journals with a higher impact factor.

Martin-Sempere et al. (2002) studied intramural and extramural collaboration. Their study found that researchers belonging to established research groups (unlike those who are affiliated to non-established groups or belong to no group) show higher scientific productivity, higher propensity to international collaboration and to participation in international projects. It was further observed that establishing a research group is advantageous for the researchers: it makes contacts and collaborations easier, encourages participation in funded projects and increases the opportunities for publication in international journals.

Mairesse (2005) studied collaboration intensity among researchers at the French Centre National de la Recherche Scientifique, in the sector of matter physics. His results confirmed that collaboration intensity between entities (towns or laboratories) is strongly and significantly correlated with productivity. The study also noted that productivity, along with the size of the scientific community, plays an important role in creating links among laboratories located in different places.

Van Raan (1998) used indicators based on citation counts to prove that the influence of international collaborations has more positive effects on the quality of the output when compared to research without collaboration. In the domain of university-industry relationships, a study by Balconi and Laboranti (2006) made an important contribution focused on the microelectronics sector. The analysis, performed on the basis of bibliometric data and information obtained through interviews, showed that the presence of university-industry collaborations is strongly correlated with the qualitative performance of jointly conducted scientific research. Other studies on public-private collaboration, rather than considering the assessment of the impact of collaboration, have aimed at identifying the industry sectors where collaboration with the academic world is most active in occurrence (Veugelers and Cassiman, 2005) and at finding potential mutual advantages of collaboration for researchers and companies (Lee, 2000; Belkhodja and Landry, 2005).

Apparently, no systematic and exhaustive study assessing the correlation between scientific performance and the various forms of collaboration intensity has yet been conducted. Explorative research seems to be favored in literature, usually involving very limited areas or individual units (departments, faculties, institutions, etc.).

## 3. Methodology and dataset

The field of investigation for this study is the whole of the Italian academic system. The data cover a total of 78 universities involved in scientific-technological disciplines. They were obtained from the ORP (Observatory on Public Research), a survey of the scientific production of all Italian public research institutions, compiled by the authors from the Scientific Citation Index of Thomson Scientific. Socio-economic and humanities disciplines are not included in the study. The analysis covers 8 of the 14 disciplinary areas (DAs) comprising the Italian academic system. These 8 areas[2], in turn, include 181 scientific-technological disciplinary sectors (SDS)[3]. On the whole, these sectors employ over 36,000 researchers, being 58% of all permanent research staff in Italian universities. The study covers the three-year period from 2001 to 2003.

---

[2] Mathematics and computer sciences, physics, chemical sciences, earth sciences, biological sciences, medical sciences, agricultural and veterinary sciences, industrial and information engineering.

[3] See http://www.miur.it/atti/2000/alladm001001_01.htm for a comprehensive list. Note that the 8 selected areas include 183 sectors, but for the two of the sectors there were no scientific publications recorded during the period 2001-2003.

Scientific publication in an international journal was used as a proxy for the production of academic research. Extramural co-authorship was used as a proxy of scientific collaboration among research organizations, with a view to identifying a possible correlation between the scientific productivity of a university and its collaboration intensity.

The total number of publications produced by the academic system during the survey period was 53,420. The following values were identified for each of the 181 sectors and each of the universities:

- publication intensity in each of its three dimensions (quantitative, qualitative and fractional);
- average quality index of scientific production;
- collaboration intensity (total, both with foreign and domestic organizations).

The following specific indicators were used:

- Output (O): total of publications authored by researchers from the university in the survey period;
- Fractional output (FO): total of the contributions made by the universities to the publications, with "contribution" defined as the reciprocal of the number of organizations with which the co-authors are affiliated;
- Scientific strength (SS): the weighted sum of the publications produced by the researchers of a university, the weights for each publication being equal to the normalized impact factor[4] of the relevant journal;
- Fractional Scientific Strength (FSS): similar to Fractional Output, but taking into account Scientific Strength;
- Productivity indicators (P, FP, QP, FQP), defined as the ratios between each of the preceding indicators and the number of university staff members in the survey period (at 31 December of the year preceding the output survey);
- Quality Index (QI): the ratio between Scientific Strength and Output, identifying the average quality of the publications produced by a university;
- Global collaboration intensity (CI), calculated as the ratio between Output and Fractional Output;
- Collaboration Intensity with foreign organizations (FCI), defined as the incidence of articles with at least one co-author affiliated to a foreign organization in the total of the publications produced by a university.
- Collaboration intensity with domestic companies (DCI), defined as the incidence of the articles with at least one co-author affiliated to a domestic enterprise in the total of the publications produced by a university.

Once distributed by sector, the data were re-aggregated by disciplinary area[5], through the steps shown in [1], [2] and [3].

---

[4] The distribution of impact factors of journals is remarkably different from one sector to another. Normalization to the sector average makes it possible to contain the distortions inherent in measurements from different sectors.

[5] This operation makes it possible to contain the bias typical of comparisons performed at high aggregation levels. Different sectors show different scientific prolificacy rates: robust comparisons are only possible through normalization of the data to the sector average and weighting by number of staff members in each sector. See Abramo et al. (2007) on this issue.

$$P_k(j) = \frac{\sum_{i=1}^{n_j} Pn_{ik} Add_{ik}}{\sum_{i=1}^{n_j} Add_{ik}} \quad [1]$$

$$IQ_k(j) = \frac{\sum_{i=1}^{n_j} IQn_{ik} Add_{ik}}{\sum_{n=1}^{n_j} Add_{ik}} \quad [2]$$

$$CI_k(j) = \frac{\sum_{i=1}^{n_j} CIn_{ik} Add_{ik}}{\sum_{n=1}^{n_j} Add_{ik}} \text{ (the same for } FCI_k(j) \text{ and } DCI_k(j)) \quad [3]$$

where:
- $P_k(j)$ = the productivity value (P, QP, FP or FQP) of university $k$ in disciplinary area $j$;
- $Pn_{ik}$ = the productivity value (P, QP, FP or FQP) of university $k$, within SDS $i$ of disciplinary area $j$, normalized to the mean of the values of all universities for SDS $i$;
- $QI_k(j)$ = Quality Index of university $k$ in disciplinary area $j$;
- $QI_{ik}$ = Quality Index of university $k$, within SDS $i$ of disciplinary area $j$, normalized to the mean of the values of all universities for SDS $i$;
- $CI_k(j)$ = Collaboration Intensity in general (with foreign organizations, $FCI_k(j)$; with private enterprises, $DCI_k(j)$) of university $k$ in disciplinary area $j$;
- $CIn_{ik}$ = Collaboration Intensity in general (with foreign organizations, $FCIn_{ik}$; with private enterprises, $DCIn_{ik}$) of university $k$, within SDS $i$ of disciplinary area $j$, normalized to the mean of the values of all universities for SDS $i$;
- $n_j$ = number of SDS included in disciplinary area $j$;
- $Add_{ik}$ = number of staff members of university $k$ affiliated to scientific-disciplinary sector $I$ of disciplinary area $j$.

In order to limit outlier-induced distortions, universities employing less than 5 staff members on average in each of the surveyed areas during the three-year period were excluded.

## 4. Findings of the study

The first outcome deriving from data elaboration refers to the association of collaboration with scientific quality. Table 1 shows cross-tabulated data for a pair of key variables: the Quality Index of research articles (measured by means of quartile of normalized journal impact factor) and type of co-authorship. Each cell reports frequency and, in brackets, the Concentration Index[6] of observations relative to specific

---

[6] Concentration Index is a measure of association between two variables based on frequencies data and varying around the neutral value of 1. Referring to the first cell of Table 1, the value of 1.33 derives from

combination of these two variables. Elaboration is aggregated at national level and data refers to scientific production of the whole Italian academic system.

| | Intramural | Extramural | | | Total |
| IF (quartile)* | | Total | With foreign organizations | With domestic enterprises | |
| --- | --- | --- | --- | --- | --- |
| 0-25 | 2,974 (1.33) | 4,507 (0.86) | 1,697 (0.70) | 221 (1.01) | 7,481 |
| 26-50 | 3,830 (1.24) | 6,443 (0.90) | 2,612 (0.79) | 313 (1.04) | 10,273 |
| 51-75 | 4,453 (1.00) | 10,369 (1.00) | 4,506 (0.94) | 419 (0.97) | 14,822 |
| 76-100 | 4,754 (0.76) | 16,090 (1.10) | 8,413 (1.25) | 607 (1.00) | 20,844 |
| Total | 16,011 | 37,409 | 17,228 | 1,560 | 53,420 |

**Table 1: Publications of Italian academic scientists by type of collaboration and quartile of normalized Impact Factor: average data 2001-2003 (concentration indexes in brackets)**
*"0" being the worst and "100" the best.*

It's evident that extramural collaboration is correlated to location of research articles in higher impact journals: a specific in-depth examination could help to understand if this is caused by a "signaling" effect on journal editors or is due to a real higher scientific impact of research output achieved in research projects performed by enlarged teams. Moreover, considering the type of extramural collaboration we found that the concentration in high impact journals is particularly remarkable for articles co-authored with foreign organizations, while there is no evidence about association of publication quality to collaboration with domestic enterprises.

**4.1 Sectorial collaboration intensity**

The aggregated at disciplinary area level, the data reveal that collaboration presents remarkable variation among disciplinary areas (Table 2). Publications involving extramural collaborations in the physical sciences account for over 95% of all publications in that area. Collaborations are only 60% of the total in the area of industrial and information engineering. The remaining areas fall between these extremes. With respect to the types of collaboration, the relative weight of the collaborating parties varies substantially from one area to another. Co-authorships with researchers from foreign organizations vary between 23.1% in medical sciences and 47.1% in the physical sciences area. Collaboration intensity with other domestic universities ($CI_{UNI}$) is rather uniform, with biological sciences (28.3%) at the top and the mathematics and computer sciences area (19.0%) at the bottom of the list. On the other hand, the incidence of collaboration with other domestic public research institutions ($CI_{DPR}$) is extremely variable. About 86% of publications in physical sciences result from this type of collaboration. This is a peculiar characteristic of this area, which includes especially large institutions (in particular, the National Nuclear Physics Institute and the National Matter Physics Institute). In the remaining areas, collaboration with other public research institutions is much less pervasive: 38% in the earth sciences area, 34.6% in medical sciences, and as little as 19% in mathematics and computer sciences. Collaboration with domestic parties, typically uninterested in scientific

---

this ratio (2,974/7,481)/(16,011/53,420) and indicates that intramural articles tend to concentrate more (+33%) in the last quartile of quality as compared with all publications.

dissemination, is especially prevalent in industrial and information engineering, involving 6.4% of publications, and in chemical sciences (3.9%). Fewer than two publications out of 1,000 in mathematics and computer sciences, and less than one in the physical sciences area, are co-authored by researchers from domestic enterprises.

| AREA | Output | CI (%) | $CI_{UNI}$ (%) | $CI_{IPR}$ (%) | FCI (%) | DCI (%) |
|---|---|---|---|---|---|---|
| Mathematics and computer sciences | 3034 | 66.3 | 19.0 | 14.0 | 37.5 | 0.6 |
| Physics | 8361 | 95.6 | 22.4 | 86.0 | 47.1 | 1.9 |
| Chemical sciences | 12347 | 69.0 | 25.7 | 26.8 | 29.1 | 3.9 |
| Earth sciences | 1706 | 78.4 | 25.0 | 38.0 | 39.2 | 2.3 |
| Biological sciences | 12770 | 71.5 | 28.3 | 29.3 | 29.6 | 2.8 |
| Medical sciences | 23766 | 66.9 | 24.2 | 34.6 | 23.1 | 2.3 |
| Agriculture and veterinary sciences | 3006 | 64.3 | 23.9 | 22.3 | 24.9 | 2.9 |
| Industrial and information engineering | 6057 | 60.6 | 20.2 | 20.8 | 24.2 | 6.4 |

**Table 2: Collaboration intensity of Italian universities as aggregated by disciplinary area and by type of collaboration; average data 2001-2003**

The variability in collaboration that emerges at the level of disciplinary areas can also be observed, and is significant, at the next lower level of aggregation, that of the scientific disciplinary sectors: Table 3 presents descriptive statistics for the distribution of collaboration intensity by disciplinary area.

| AREA | Mathematics and computer sciences | Physics | Chemical sciences | Earth sciences | Biological sciences | Medical sciences | Agricultural and veterinary sciences | Industrial and information engineering |
|---|---|---|---|---|---|---|---|---|
| Number of SDS | 10 | 8 | 12 | 12 | 19 | 49 | 30 | 41 |
| Average (%) | 68.2 | 92.7 | 66.9 | 77.3 | 71.5 | 65.0 | 63.9 | 57.5 |
| Median (%) | 70.0 | 94.5 | 66.6 | 79.7 | 72.8 | 65.0 | 63.3 | 54.9 |
| Min (%) | 55.6 | 83.3 | 39.8 | 62.2 | 60.4 | 39.0 | 33.3 | 31.8 |
| Max (%) | 77.1 | 97.4 | 93.3 | 87.1 | 83.1 | 100.0 | 87.3 | 80.0 |
| Std.dev (%) | 6.2 | 5.1 | 13.0 | 8.4 | 6.2 | 13.2 | 10.6 | 10.9 |
| Coeff variation | 0.091 | 0.056 | 0.195 | 0.109 | 0.087 | 0.203 | 0.166 | 0.190 |

**Table 3: Sectors with highest collaboration intensity with foreign parties, for each disciplinary area (in brackets: percentage of incidence of that sector output on area total**

Only 3 of the 8 areas investigated show a variation coefficient of these distributions that is lower than 0.1 (mathematics and computer sciences, physical sciences and biological sciences)[7]. In the medical sciences area, the maximum-minimum difference is 61%, as compared to 54% in agricultural and veterinary sciences and chemical sciences.

Sector-specific characteristics also become evident and significant if co-authorships with researchers affiliated to foreign organizations or domestic enterprises are considered. For the sake of brevity, the data showing such characteristics have been summarized for this paper. Table 4 and 5 indicate, for each area, the sector with the highest degree of collaboration with foreign and domestic parties, respectively.

---
[7] The coefficient of variation is a normalized measure of dispersion of a distribution. It is defined as the ratio of the standard deviation to the mean and its dimensionless nature renders comparable different data sets with wildly different means. The larger is its value the broader is the data dispersion of the distribution.

The degree of internationalization of academic scientific production (Table 4) peaks at 65% in sectors FIS/05 (astronomy and astrophysics) and AGR/05 (forest management and silviculture). The latter sector is in fact extremely small in size, with only 55 publications over the three-year period. The cases of the genetics (BIO/18) and medical genetics (MED/03) sectors are particularly interesting: 40% of the 500 publications surveyed over the period in the two sectors were co-authored by foreign researchers. General and inorganic chemistry (CHIM/03) also shows very significant collaboration intensity with foreign parties, accounting for more than 38% of the 3,000 publications surveyed.

| AREA | SDS | FCI (%) | Output |
|---|---|---|---|
| Mathematical and computer sciences | MAT/01 - Mathematical logic | 57.1 | 35 (1.2%) |
| Physics | FIS/05 - Astronomy and astrophysics | 64.5 | 887 (10.6%) |
| Chemical sciences | CHIM/03 - General and inorganic chemistry | 38.2 | 2857 (23.1%) |
| Earth sciences | GEO/01 - Paleontology and paleoecology | 56.5 | 92 (5.4%) |
| Biological sciences | BIO/18 - Genetics | 40.5 | 538 (4.2%) |
| Medical sciences | MED/03 - Medical genetics | 39.9 | 469 (2.0%) |
| Agricultural and veterinary sciences | AGR/05 - Forest management and silviculture | 65.5 | 55 (1.8%) |
| Industrial and information engineering | ING-IND/19 - Nuclear plants | 41.3 | 80 (1.3%) |

**Table 4: Sectors with highest collaboration intensity with domestic enterprises, for each disciplinary area (in brackets: percentage of incidence of sector output on area total)**

Co-authorships with researchers from domestic enterprises (Table 5) are substantially less frequent: Electronics, with 12.6% of the 905 surveyed publications, and industrial chemistry, with 10.9%, are the best-performing SDS. On the other hand, physical sciences and mathematical sciences, being typical basic research areas, seldom use collaborations with domestic parties, and account for, at best, less than 3% of the total.

| AREA | SDS | DCI (%) | Output |
|---|---|---|---|
| Mathematics and computer sciences | MAT/07 - Mathematical physics | 1.5 | 548 (18.1%) |
| Physics | FIS/07 - Applied physics | 2.7 | 699 (8.4%) |
| Chemical sciences | CHIM/04 - Industrial chemistry | 10.9 | 713 (5.8%) |
| Earth sciences | GEO/02 - Stratigraphic and sedimentary geology | 7.4 | 149 (8.7%) |
| Biological sciences | BIO/15 - Pharmaceutical biology | 5.2 | 229 (1.8%) |
| Medical sciences | MED/24 – Urology | 6.2 | 243 (1.0%) |
| Agricultural and veterinary sciences | AGR/19 - Animal husbandry | 5.9 | 185 (6.2%) |
| Industrial and information engineering | ING-INF/01 – Electronics | 12.6 | 905 (14.9%) |

**Table 5: Sectors with higher collaboration intensity with domestic enterprises for each disciplinary area (in brackets: percentage of incidence of sector output on the total of the area)**

## 4.2 Collaboration and scientific performance

This section first provides a sectorial analysis of the types and intensity of research collaborations, and then attempts to identify an answer for our primary question: is there a correlation between a university's collaboration intensity and its scientific performance? In other words: are the universities whose researchers are more actively involved in extramural collaborations the most productive, and if so, to what extent? Table 6 presents the statistics regarding the correlation analysis between the relevant performance indicators and overall collaboration intensity at university level[8]. No simple rule can be derived from the data, as different correlation degrees emerge in different areas.

|     | Statistics   | Mathematics and computer sciences | Physics | Chemical sciences | Earth sciences | Biological sciences | Medical sciences | Agricultural and veterinary sciences | Industrial and information engineering |
|-----|--------------|-----|------|-------|------|------|------|------|------|
| P   | *Correlation.* | 0.22 | 0.11 | 0.06 | 0.36 | 0.39 | 0.09 | 0.54 | 0.68 |
|     | $\beta$      | 0.52 | 0.12 | 0.08 | 0.61 | 0.80 | 0.23 | 0.59 | 0.99 |
|     | $R^2$        | 0.05 | 0.01 | 0.00 | 0.13 | 0.15 | 0.01 | 0.30 | 0.47 |
| FP  | *Correlation.* | 0.06 | -0.25 | -0.42 | 0.13 | 0.03 | -0.13 | 0.25 | 0.44 |
|     | $\beta$      | 0.10 | -0.14 | -0.41 | 0.12 | 0.04 | -0.26 | 0.25 | 0.51 |
|     | $R^2$        | 0.00 | 0.06 | 0.17 | 0.02 | 0.00 | 0.02 | 0.06 | 0.20 |
| QP  | *Correlation.* | 0.32 | 0.18 | 0.05 | 0.31 | 0.42 | 0.11 | 0.52 | 0.68 |
|     | $\beta$      | 0.91 | 0.21 | 0.08 | 0.50 | 1.07 | 0.30 | 0.63 | 1.04 |
|     | $R^2$        | 0.11 | 0.03 | 0.00 | 0.10 | 0.18 | 0.01 | 0.27 | 0.47 |
| FQP | *Correlation.* | 0.20 | -0.20 | -0.37 | -0.01 | 0.13 | -0.08 | 0.30 | 0.42 |
|     | $\beta$      | 0.38 | -0.10 | -0.39 | -0.01 | 0.19 | -0.19 | 0.32 | 0.50 |
|     | $R^2$        | 0.04 | 0.04 | 0.14 | 0.00 | 0.02 | 0.01 | 0.09 | 0.17 |
| QI  | *Correlation.* | 0.54 | 0.35 | 0.13 | 0.49 | 0.45 | 0.30 | 0.73 | 0.70 |
|     | $\beta$      | 0.48 | 0.11 | 0.06 | 0.30 | 0.26 | 0.21 | 0.46 | 0.44 |
|     | $R^2$        | 0.30 | 0.12 | 0.02 | 0.24 | 0.20 | 0.09 | 0.53 | 0.48 |

**Table 6: Statistics regarding the association of collaboration intensity (CI) with the performance of universities, by disciplinary area**

Industrial and information engineering is the only area in which a strong correlation emerges between collaboration intensity and scientific performance in all its dimensions. The statistical analyses give similar results for the areas of biological sciences and agricultural and veterinary sciences. Collaboration intensity appears to be significantly correlated to productivity (P and QP) as well as to average quality of scientific production. Mathematics and computer sciences show a significant correlation in the average quality of scientific production (QI) only, which proves that the results of research involving extramural collaboration are published in more prestigious journals than those involving intramural collaboration.

---

[8] Hereinafter we'll make an extensive use of association analysis by means of:
- ✓ correlation coefficient: it measures the strength and direction of a linear relationship between two variables on a 0-1 scale, where "0" represents no association and "1" perfect association.
- ✓ regression statistics: they measure the relationship between an independent variable (X) and a dependent variable (Y). In our case we show the coefficient ($\beta$) of such relationship and the coefficient of determination ($R^2$) which measures (on a 0-1 scale) the variability of Y explained by X.

In general, the average quality of scientific production is the variable that most often correlates positively to the collaboration intensity of universities. The area of chemical sciences provides an exception. Interestingly, the indicators of fractional productivity for this area actually seem to be negatively correlated with collaboration intensity. What this probably means is that, since outward-oriented universities tend to share their scientific output with third parties, they are negatively affected in terms of contribution.

**4.3 The effect of international collaboration**

Moving on to our second question, Table 7 maps the statistics regarding the correlation between performance indicators at university level and collaboration intensity with foreign organizations, for each of the disciplinary areas covered.

For the case of the physical sciences area, all the performance indicators appear to be significantly correlated with collaboration intensity with foreign organizations, but the correlation is weak in terms of quality. The same degree of significance, excluding QI, is found for the chemical sciences area. The correlation is significant with respect to productivity and qualitative productivity in the earth sciences and industrial and information engineering areas. On the other hand, only indicators based also on quality (QP, FQP, QI) are associated with the internationalization of scientific production in biological sciences and medical sciences. In the latter area, correlation is very strong (0.77) for quality index (QI) suggesting that international co-authorship is remarkably associated with higher quality of publication location of search results.

|  | Statistics | Mathematics and computer sciences | Physics | Chemical sciences | Earth sciences | Biological sciences | Medical sciences | Agricultural and veterinary sciences | Industrial and information engineering |
|---|---|---|---|---|---|---|---|---|---|
| P | *Correlation.* | 0.16 | 0.50 | 0.39 | 0.46 | 0.26 | 0.23 | 0.33 | 0.44 |
|  | $\beta$ | 0.29 | 0.87 | 0.45 | 0.76 | 0.39 | 0.47 | 0.25 | 0.64 |
|  | $R^2$ | 0.03 | 0.25 | 0.15 | 0.21 | 0.07 | 0.05 | 0.11 | 0.19 |
| FP | *Correlation.* | 0.08 | 0.34 | 0.53 | 0.27 | 0.11 | 0.23 | 0.11 | 0.24 |
|  | $\beta$ | 0.10 | 0.28 | 0.45 | 0.24 | 0.10 | 0.41 | 0.08 | 0.28 |
|  | $R^2$ | 0.01 | 0.11 | 0.28 | 0.07 | 0.01 | 0.06 | 0.01 | 0.06 |
| QP | *Correlation.* | 0.27 | 0.50 | 0.37 | 0.42 | 0.34 | 0.31 | 0.35 | 0.46 |
|  | $\beta$ | 0.57 | 0.88 | 0.47 | 0.64 | 0.37 | 0.72 | 0.29 | 0.71 |
|  | $R^2$ | 0.07 | 0.25 | 0.14 | 0.18 | 0.22 | 0.10 | 0.12 | 0.21 |
| FQP | *Correlation.* | 0.21 | 0.35 | 0.49 | 0.13 | 0.23 | 0.30 | 0.18 | 0.23 |
|  | $\beta$ | 0.30 | 0.27 | 0.44 | 0.14 | 0.25 | 0.57 | 0.13 | 0.29 |
|  | $R^2$ | 0.04 | 0.12 | 0.24 | 0.02 | 0.05 | 0.09 | 0.03 | 0.06 |
| QI | *Correlation.* | 0.44 | 0.28 | 0.07 | 0.42 | 0.48 | 0.77 | 0.57 | 0.42 |
|  | $\beta$ | 0.29 | 0.13 | 0.03 | 0.25 | 0.31 | 0.45 | 0.25 | 0.28 |
|  | $R^2$ | 0.19 | 0.08 | 0.01 | 0.17 | 0.23 | 0.59 | 0.32 | 0.18 |

**Table 7: Statistics regarding the association of collaboration intensity with foreign organizations (FCI) with the performance of universities by disciplinary area**

Unlike all other areas, mathematical sciences and agricultural and veterinary sciences show a significant correlation only with the average quality of scientific production.

In keeping with the results of the preceding analysis, this dimension of performance appears to be especially sensitive to the degree of internationalization of the scientific portfolios of universities in all the areas covered, except for chemical sciences.

**4.4 Impact from collaboration with domestic enterprises**

The analysis presented in the two preceding sections clearly indicates that the degree of correlation between extramural collaboration intensity and a university's research performance varies according to performance indicators and according to area. This final section will focus on our third research question. The aim is to verify whether the universities with the best scientific performances are those with the highest collaboration intensity with enterprises. Table 8 presents the statistics regarding the correlation between performance indicators, as measured at university level, and collaboration intensity with domestic enterprises.

|  | Statistics | Mathematics and computer sciences | Physics | Chemical sciences | Earth sciences | Biological sciences | Medical sciences | Agricultural and veterinary sciences | Industrial and information engineering |
|---|---|---|---|---|---|---|---|---|---|
| P | *Correlation* | -0.04 | 0.31 | 0.14 | 0.44 | 0.48 | 0.15 | 0.17 | 0.16 |
|  | $\beta$ | -0.02 | 0.15 | 0.07 | 0.25 | 0.31 | 0.14 | 0.05 | 0.13 |
|  | $R^2$ | 0.00 | 0.10 | 0.02 | 0.20 | 0.23 | 0.02 | 0.03 | 0.03 |
| FP | *Correlation* | 0.01 | 0.41 | 0.06 | 0.31 | 0.47 | 0.13 | 0.15 | 0.15 |
|  | $\beta$ | 0.00 | 0.09 | 0.02 | 0.09 | 0.19 | 0.10 | 0.04 | 0.10 |
|  | $R^2$ | 0.00 | 0.17 | 0.00 | 0.10 | 0.22 | 0.02 | 0.02 | 0.02 |
| QP | *Correlation.* | -0.06 | 0.25 | 0.20 | 0.38 | 0.47 | 0.15 | 0.17 | 0.22 |
|  | $\beta$ | -0.04 | 0.12 | 0.11 | 0.20 | 0.37 | 0.16 | 0.06 | 0.19 |
|  | $R^2$ | 0.00 | 0.06 | 0.04 | 0.14 | 0.22 | 0.02 | 0.03 | 0.05 |
| FQP | *Correlation* | -0.03 | 0.38 | 0.12 | 0.22 | 0.48 | 0.13 | 0.19 | 0.21 |
|  | $\beta$ | -0.01 | 0.08 | 0.05 | 0.08 | 0.22 | 0.11 | 0.06 | 0.14 |
|  | $R^2$ | 0.00 | 0.14 | 0.02 | 0.05 | 0.23 | 0.02 | 0.04 | 0.04 |
| QI | *Correlation.* | 0.13 | 0.12 | 0.40 | 0.32 | 0.21 | -0.07 | -0.04 | 0.32 |
|  | $\beta$ | 0.02 | 0.02 | 0.06 | 0.07 | 0.04 | -0.02 | -0.01 | 0.11 |
|  | $R^2$ | 0.02 | 0.01 | 0.16 | 0.10 | 0.04 | 0.01 | 0.00 | 0.10 |

**Table 8: Statistics regarding the association of collaboration intensity with domestic enterprises (DCI) with the performance of universities, by disciplinary area**

The data in the table clearly indicate that in 3 of the 8 areas covered (mathematics and computer sciences, medical sciences, agricultural and veterinary sciences), collaboration with domestic parties is in no way related to the universities' performances. In two other areas (industrial and information engineering, chemical sciences), the sole association involves the average quality, with a weak correlation. As shown in Table 2, these are the areas in which university-enterprise collaborations are most frequent (6.4% in the former, 3.9% in the latter).

In biological sciences, all the productivity indicators show a significant correlation with the frequency of public-private co-authorships. The same is also true, at least for some performance dimensions, in the physical sciences area (where, however, public-private collaborations have a lower incidence on the total of publications surveyed) and in earth sciences.

**Conclusions**

Our study has revealed some salient characteristics of collaboration in academic research. It becomes apparent that collaboration intensity is not uniform, but rather depends on the specific scientific field considered. Sectors which are by definition interdisciplinary, with production of cross-sector knowledge in different research domains, certainly have a stronger tendency to use collaboration than more "vertical" sectors, where research tends to be relatively more intra-mural. Basic research clearly appears to stimulate collaboration with foreign organizations more than applied research. In basic research, more complex phenomena are analyzed and greater instrumental resources are needed to obtain a significant advance in knowledge, all of which calls for assets and competencies that are typically delocalized. On the other hand, collaboration with domestic parties is rather low due to the reluctance of private enterprises to share proprietary knowledge, and is most frequent in sectors where research is mainly application-oriented.

Our analysis cannot give a straightforward answer to our first research question. What emerges is that the correlation degree between productivity and extramural collaboration intensity varies substantially among different areas. The only area showing a strong correlation for all the relevant dimensions of performance is that of industrial and information engineering. In general, the average quality of scientific production is the variable that most often correlates positively to the extramural collaboration intensity of universities.

The presence of sectoral peculiarities is confirmed by the analysis of international collaborations. It is again the average quality of scientific production that presents a positive correlation with the degree of internationalization of the scientific production of universities. In some areas, however, a significant correlation with the productivity indicators of universities is also evident. This is especially true for the physical sciences, but also for the areas of chemical sciences, earth sciences and industrial and information engineering.

Collaborations with domestic parties show a strong correlation with productivity in the area of biological sciences, and to a lesser degree, in physical sciences and earth sciences. Possible factors determining such a degree of correlation with research productivity, especially in the sectors where it is most evident, include the following:
  i)   complementary assets owned by private parties are critical to achieving significant advances in knowledge;
  ii)  companies resort to collaboration with prominent scientists in order to build support around their research activities among stakeholders;
  iii) companies operating in these sectors have developed a better ability to select the best researchers as their scientific partners.

The authors expect to be able to test these hypotheses empirically in the future.

In general, there is no clear evidence that correlation exists between the resort to extramural collaboration and the overall performance of a research institution, even though there are clear differences with regard to the several types of collaboration (general, international, public-private) and the peculiarities of the specific fields of scientific research. Even significant differences in efficiency among academic research organizations do not necessarily reflect a different degree of extramural collaboration.

However, there are sound and reasonable arguments remaining to support policy measures in favor of networking and collaborations among research groups. In particular, one of those arguments is that networking and collaborations help disseminate knowledge and research results more rapidly and pervasively.

**References**

Abramo G., D'Angelo C.A., Pugini F., (2007). The Measurement of Italian Universities' Research Productivity by a non Parametric-Bibliometric Methodology, *Scientometrics*, 76, 2.

Adams S.J.D., Black G.C., Clemmons J.R., Paula E., Stephan P. E., (2005). Scientific teams and institutional collaborations: Evidence from U.S. universities, 1981–1999, *Research Policy* 34, 3, 259-285.

Balconi M., Laboranti A., (2006).University–industry interactions in applied research: The case of microelectronics. *Research Policy*, 35, 1616–1630.

Barjak F., Robinson S., (2007). International collaboration, mobility and team diversity in the life sciences: impact on research performance, *Social Geography Discussion*, 3, 121-157.

Belkhodja O., Landry R., (2005). The Triple Helix collaboration: Why do researchers collaborate with industry and the government? What are the factors influencing the perceived barriers?, *5th Triple Helix Conference*, Turin, Italy, May 18th-21st 2005.

Bordons M., Gómez I., Fernández M. T., Zulueta M. A. Méndez A.,(1996). Local, Domestic and International Scientific Collaboration in Biomedical Research, *Scientometrics*, 37, 2, 279-295.

Commission of the European Communities, *COM 353 final,* (2004). Science and technology, the key to Europe's future - Guidelines for future European Union policy to support research.

Commission of the European Communities, *COM 161*, (2007). Libro Verde - Nuove prospettive per lo Spazio europeo della ricerca.

Dundar H., Lewis D. R., (1998). Determinants of research productivity in higher education, *Research in Higher Education*, 39, 6, 607-631.

Fox, M.F., (1983). Publication productivity among scientists: a critical review, *Social Studies of Science* 13, 285-305.

Fox M.F., (1992a). Research, teaching, and publication productivity: mutuality versus competition in academia. *Sociology of Education* 65, 293-305.

Fox M.F., (1992b). Research productivity and environmental context, in Whiston, T.G., and Geiger, R.L. (eds.), *Research and Higher Education*. Buckingham: SRHE and Open University Press.

Katz J.S., Martin B.R., (1997). What is research collaboration?, *Research Policy,* 26, 1-18.

Landry R., Traore N., Godin B., (1996). An econometric analysis of the effect of collaboration on academic research productivity. *Higher Education*, 32, 283-301.

Lee S., (2000). The Sustainability of University-Industry Research Collaboration: An Empirical Assessment, *Journal of Technology Transfer,* 25, 2, 111-133.

Lee S., Bozeman B., (2005). The Impact of Research Collaboration on Scientific Productivity, *Social Studies of Science*, 35, 5, 673-702.

Mairesse J.; Turner L., (2005). Measurement and explanation of the intensity of co-publication in scientific research: an analysis at the laboratory level, *Nber Working Paper Series*, w11172.

Martin-Sempere M. J., Rey-Rocha J., Garzon-Garcia B., (2002). The effect of team consolidation on research collaboration and performance of scientists. Case study of Spanish University researchers in Geology, *Scientometrics*, 55, 3, 377-394.

Ramsden P., (1994). Describing and explaining research productivity, *Higher Education,* 28, 207-226

Van Raan A.F.J., (1998). The influence of international collaboration of the impact of the research results, *Scientometrics*, 42, 3, 423-428.

Veugelers R., Cassiman B., (2005). R&D cooperation between firms and universities. Some empirical evidence from Belgian manufacturing, *International Journal of Industrial Organization*, 23, 355– 379.